\documentclass{PoS}
\def \half{{\textstyle{\frac{1}{2}}}}

\title{Updates of PDFs in the MMHT framework}

\ShortTitle{MMHT -- DIS2019}

\author{\speaker{R.S.~Thorne} \\
        Department of Physics and Astronomy, \\
        University College London, WC1E 6BT, UK\\
        E-mail: \email{robert.thorne@ucl.ac.uk}}

\author{S.~Bailey  \\
        Rudolf Peierls Centre, Beecroft Building,\\  
        Parks Road, Oxford, OX1 3PU, UK\\
        E-mail: \email{shaun.bailey@physics.ox.ac.uk}}

\author{T.~Cridge \\
        Department of Physics and Astronomy, \\
        University College London, WC1E 6BT, UK\\
        E-mail: \email{t.cridge@ucl.ac.uk}}

\author{L.A.~Harland-Lang\\
        Rudolf Peierls Centre, Beecroft Building,\\  
        Parks Road, Oxford, OX1 3PU, UK\\
        E-mail: \email{lucian.harland-lang@physics.oxford.ac.uk}}

\author{A.D.~Martin\\
        Institute for Particle Physics Phenomenology,\\
        University of Durham, DH1 3LE, UK \\ 
        E-mail: \email{A.D.Martin@durham.ac.uk}}

\author{R.~Nathvani \\
        Department of Physics and Astronomy, \\
        University College London, WC1E 6BT, UK\\
        E-mail: \email{ricky.nathvani.15@ucl.ac.uk}}

\abstract{We summarise recent developments in the path towards the 
``MMHT19'' parton distribution functions. We concentrate on the extraction of 
the strange quark upon the improvement of theoretical calculations for 
NNLO charged current cross sections; the effect of an extension of our 
parameterisation; and the role of correlated uncertainties in some data sets 
which prove difficult to fit.}

\FullConference{XXVII International Workshop on Deep-Inelastic Scattering and Related Subjects - DIS2019\\
		8-12 April, 2019\\
		Torino, Italy}

\begin{document}

The MMHT2014 PDFs \cite{Harland-Lang:2014zoa}
were the last major update in the MRST/MSTW/MMHT family of PDFs.
They included a variety of LHC data in their determination, but at that point
the constraint on the PDFs was almost entirely from older fixed target and HERA
DIS data, and some Drell-Yan and jet data from Tevatron experiments. 
Soon after the publication of the MMHT2014 PDFs we 
studied the effect of including the final HERA total cross 
section measurements \cite{Abramowicz:2015mha}, noting 
only minor changes in the central values and uncertainties 
\cite{Harland-Lang:2016yfn}. The inclusion of a wide variety of new LHC data 
was considered in  \cite{Thorne:2017aoa}. Most new data was fit well, and 
produced relatively small changes in both the central values and the 
uncertainties of
the PDFs, most notably in the decomposition into flavours and into valence 
and sea quarks. A more significant change was induced when the ATLAS $W,Z$ 
data in \cite{Aaboud:2016btc} were included. A relatively good fit could 
be achieved ($\sim 110/61$), but this required a modification of the small-$x$ 
valence quarks and, in particular, an increase in the strange quark. This
latter change was less pronounced than that found by ATLAS in
 \cite{Aaboud:2016btc}, probably
due to the inclusion of fixed target charged current DIS dimuon production 
data \cite{Goncharov:2001qe}, which is the traditional type of constraint 
on the strange quark in global PDF fits and prefers a lower value than the 
ATLAS data. 

\vspace{0.0cm}
\begin{table}[h]
\scriptsize
\begin{center}
\begin{tabular}{|l|c|c|c|c|}
\hline
             & BR$(c \to \mu)$& CCFR/NuTeV $\chi^2$  &   ATLAS $W,Z$ $\chi^2$ &
  Total $\chi^2$         \\
\hline
MMHT+HERAII                                 &  0.090  & 120.5 &       & 3526.3  \\
MMHT+HERAII (NNLO dimuon )                  &  0.102  & 122.7 &       & 3527.3    \\
MMHT+HERAII (NNLO VFNS dimuon)              &  0.101  & 123.9 &       & 3531.3     \\
MMHT+HERAII+ATLAS($W,Z$)                    &  0.073  & 127.3 & 108.6 & 3684.7  \\
MMHT+HERAII+ATLAS($W,Z$) (NNLO dimuon )     &  0.084  & 137.8 & 106.8 & 3688.4    \\
MMHT+HERAII+ATLAS($W,Z$) (NNLO VFNS dimuon) &  0.086  & 137.0 & 106.8 & 3688.5     \\
\hline                                                        
$N_{pts}$      &     & 126  & 61  & 3337  \\
\hline
    \end{tabular}
\caption{The branching ratio for charm mesons to muons, and the $\chi^2$ values 
for fit variants before and after the inclusion of the data in 
\cite{Aaboud:2016btc} and the correction to the NNLO dimuon cross sections.} 
\vspace{-0.4cm}
\end{center}
\end{table}

However, the full NNLO corrections to heavy flavour charged current DIS
have only recently been calculated \cite{Berger:2016inr}, and all PDF analyses
so far have used either NLO cross sections, or NNLO approximations 
(as in MMHT), in 
NNLO PDF extractions. The NNLO corrections are of order $-10\%$ in the region 
$x=0.01$ relevant for the main strange quark sensitivity of the  
ATLAS data. This implies a larger strange quark may be needed to best fit 
the dimuon data and better potential compatibility with the ATLAS data. 
We have included this full NNLO calculation in a complete PDF fit for the 
first time, and modified our variable flavour number scheme (VFNS) for charged 
currents to include the calculation, which is for fixed flavour number (this 
required some minor corrections to the details of our charged current VFNS). 
The results of the fit quality are shown in Table 1. 
We note that when the ATLAS $W,Z$ data are not included the inclusion of the 
correct NNLO makes little difference to fit quality, though it does raise the
preferred value of the branching ratio for $c \to \mu$ in the dimuon fit a 
little, consistent with a smaller charm cross section. The input value of 
the branching ratio is $0.092 \pm 10 \%$, very consistent with these fits.
When the ATLAS $W,Z$ data are added using the standard MMHT treatment of 
dimuons the fit to the dimuon data deteriorates a little, but the branching 
ratio also has to take a very low value of 0.073 to accommodate the raised 
strange quark. With the full NNLO corrections the dimuon fit actually worsens, 
but that to ATLAS $W,Z$ data improves slightly and the global $\chi^2$ changes 
by less than 5 units. The branching ratio increases back to 0.086, however, 
demonstrating more genuine compatibility between the fit to dimuon data and to 
ATLAS data. The effect on $s + \bar s$ is shown in Fig. 1. With both
the old MMHT treatment and the corrected NNLO cross section inclusion of the 
ATLAS $W,Z$ data raises the strange quark for $x$ above $0.001$ and reduces the 
uncertainty. However, the rise is a little less significant when the full NNLO
is used, with the updated strange central value straying outside the one sigma
error band of the fit without ATLAS $W,Z$ data only marginally near $x=0.02$.  

\vspace{-0.3cm}
\begin{figure}[]
\centerline{\includegraphics[width=0.48\textwidth]{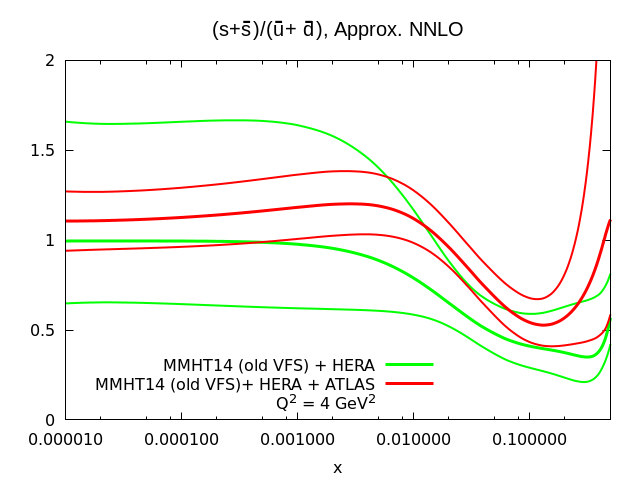}
\includegraphics[width=0.48\textwidth]{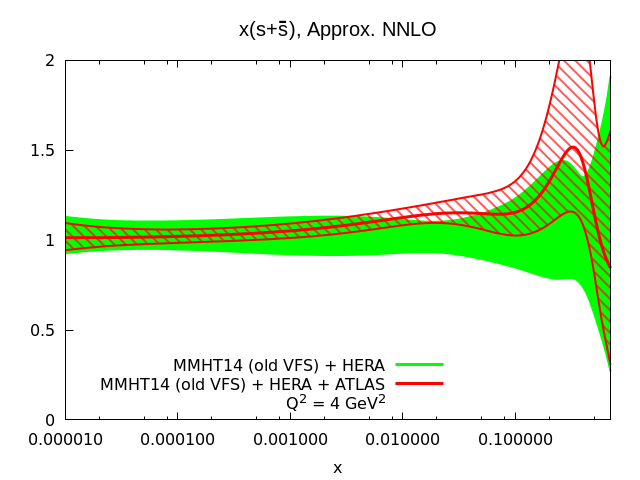}}
\centerline{\includegraphics[width=0.48\textwidth]{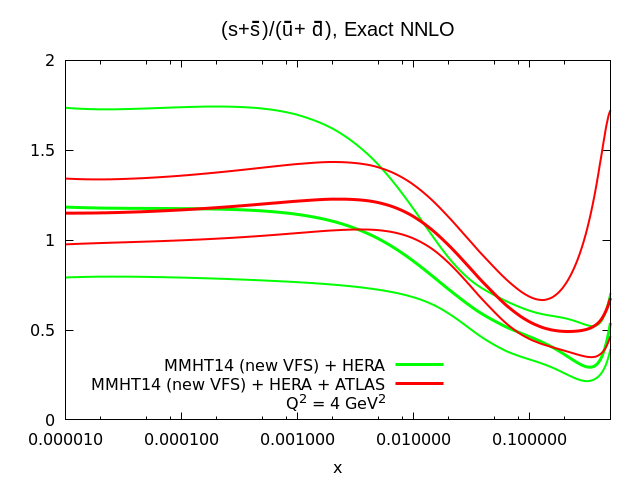}
\includegraphics[width=0.48\textwidth]{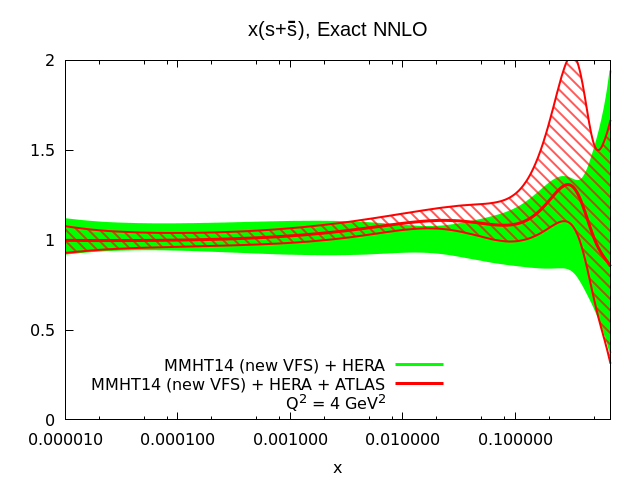}}
\vspace{-0.2cm}
\caption{The strange over the light quark average and change in strange 
on addition of ATLAS $W,Z$ data without full NNLO as in MMHT2014 (top),
and with full NNLO and the updated VFNS (bottom).}
\vspace{-0.2cm}
\label{Fig1} 
\end{figure}

With the inclusion of more constraining LHC data
we examine the introduction of more free parameters in our functions for 
the input PDFs. In \cite{Martin:2012da} we changed our previous parameterisation 
to one with most PDFs parameterised as $A(1-x)^{\eta} x^{\delta}
(1+\sum_{i=1}^n a_i T_i(1-2x^{\half}))$, where $T_i(1-2x^{\half}))$ 
are Chebyshev polynomials. We also examined how many were needed to fit data 
of a given precision. At the time it was concluded that 4 (at most) was 
sufficient. However, for $(\bar d -\bar u)(x,Q_0^2)$ 
by default we still used the old input form with 
only 4 parameters in total. We now try 
putting this distribution on an equal footing with the other PDFs by using 
Chebyshev polynomials, i.e. 

\vspace{0.2cm}
$(\bar d -\bar u)(x,Q_0^2) = 
A(1-x)^{\eta_{sea}+2} x^{\delta}(1+\gamma x + \Delta x^2) \to
A(1-x)^{\eta_{sea}+2} x^{\delta} (1+\sum_{i=1}^{m} a_i T_i(1-2x^{\half})),$
\vspace{0.2cm}

\noindent where we first choose $m=4$. This improves the global fit by about 10 
units and eases tension between 
ATLAS $W,Z$ data and E866 Drell Yan ratio data \cite{Towell:2001nh}.  

However, from the studies in  \cite{Martin:2012da} it was clear that once PDF
precision of greater than $1\%$ was possible 5 or 6 polynomials would be  
required. Hence we extend the parameterisation for $u_V, d_V, {\rm sea},
(\bar d -\bar u)$ and
$(s + \bar s)$ sequentially to the form with $n=6$ 
(though for $(s + \bar s)$ the  small-$x$ power and two coefficients remain 
tied to the sea). The gluon 
parameterisation is of a different form, with two separate terms, but here we 
increase the free parameters from 7 to 9. Compared to MMHT2014, this increases 
our number of free parameters in total from 36 to 50. The main improvements 
in the fit $\chi^2$ (to essentially the same data as in \cite{Thorne:2017aoa}) 
are after the extension to 6 polynomials of 
$(\bar d -\bar u)(x,Q_0^2)$, from additionally
extending $d_V(x,Q_0^2)$ ($u_V(x,Q_0^2)$ is not significant), 
and $g(x,Q_0^2)$ (${\rm sea}(x,Q_0^2)$ and $s^+(x,Q_0^2)$
are not significant). The improvements to individual data sets are shown in 
Table 2.  There is a reduction in tension between DY ratio data and LHC data, 
and an improvement in LHC lepton asymmetry data, and some 
fixed target deuteron data.
The improvement due to the gluon is spread over data sets and due only 
partially to HERA data.

\begin{table}[h]
\begin{center}
\begin{tabular}{|l|c|c|c|}
\hline
  Data set           & $-\Delta \chi^2 \quad (\bar d -\bar u)$ &$-\Delta \chi^2\quad (\bar d -\bar u), d_V$  &   $-\Delta \chi^2$ All \\
\hline
Total                    & 17.6  & 34.0 &  48.9  \\
\hline
BCDMS $F_2^p$       & -4.6  & -3.3 & -2.7    \\
BCDMS $F_2^d$       & -2.7  & 4.9 & 8.5    \\
NMC $F_2^n/F_2^p$   &  6.5  & 6.1 & 6.0  \\
NuTeV $F_3^{N}$    & -0.3  & 1.7 & 3.2    \\
E866 $\sigma(pd)/\sigma(pp)$  &  8.2 & 10.1 & 11.0    \\
NuTeV dimuon       & 0.7  & 1.0 & 3.0    \\
HERA I+II $\sigma(e^+p)~920$~GeV       &  1.1 & 1.7 & 4.6     \\
CMS $pp\to l^+l^-$      & 0.7  & 1.8 & 3.1    \\
D0 $\sigma(e^+) - \sigma(e^-)$  & -1.2&  -3.4 & -1.4    \\
CMS 8~TeV$ \sigma(l^+) - \sigma(l^-)$  & 4.4  &5.0 & 4.6    \\
ATLAS 7~TeV $W,Z$  & -0.5  &2.2 & 4.3    \\
CMS 7~TeV jets  & -0.5  &0.2 & 3.2    \\
\hline
    \end{tabular}
\caption{The $\chi^2$ improvement for various data sets when extra parameters 
are added.}
\vspace{-0.4cm}
\end{center}
\end{table}

When determining the PDF uncertainties we go from 25 eigenvector pairs in 
MMHT2014 to 30, one extra parameter for each PDF
other than the light sea (and $s-\bar s$). 20 parameters are fixed at their 
best fit values -- any extra eigenvectors would be highly non-quadratic 
and lead to very little extra uncertainty. The mean tolerance $T=3.31$, 
similar to MMHT2014. 27 eigenvector directions are constrained primarily by 
LHC data sets,
largely $7$~TeV ATLAS $W,Z$ data and CMS $W$ data.
E866 Drell Yan asymmetry is vital for constraining $\bar d -\bar u$. 
Tevatron data of various types are the 
primary constraint for 8 eigenvectors, and
fixed target DIS data  still constrains 12
eigenvectors (mainly high-$x$ quarks).
Hence, a fully global fit is necessary for full constraint. Some of the 
resulting PDFs are shown in Figs. 2 and 3. There is a significant change
in the shape of the $d_V$ distribution. 
The uncertainty at both large and small $x$ increases due to the 
extra freedom. $d_V$ interplays with $(\bar d - \bar u)$, also shown in 
Fig. 2, and the latter 
becomes flatter over the range $x=0.01-0.1$. The gluon, shown
in Fig.3, stays well within previous uncertainties, except at very high $x$ 
where the uncertainty is huge and larger with the new parameterisation.
$\bar u$, also shown in Fig. 3, is generally a little smaller, due to the 
increase in $(s + \bar s)$ already discussed, and again has increased 
uncertainty at $x>0.6$. We note that despite more freedom in $(\bar d -\bar u)$
the parameterisation still leads to a very quick approach to 0 at low $x$. This 
is likely unphysical, and we have recently tried parameterising 
$\bar d/\bar u$ instead, using the same number of parameters. Results are  
preliminary, but a small decrease in $\chi^2$ occurs, with a greater
uncertainty as $x \to 0$ (though $\bar d /\bar u \to 1$ to very good accuracy), 
and with a ratio of $\bar d /\bar u < 1$ at low $x$, but only at about 
one sigma level.   
       
\vspace{-0.3cm}
\begin{figure}[]
\centerline{\includegraphics[width=0.5\textwidth]{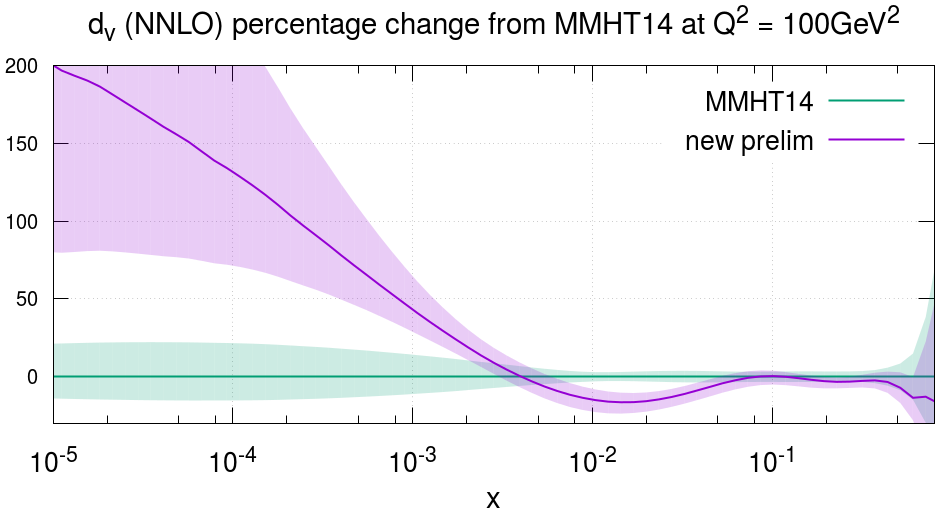}
\includegraphics[width=0.5\textwidth]{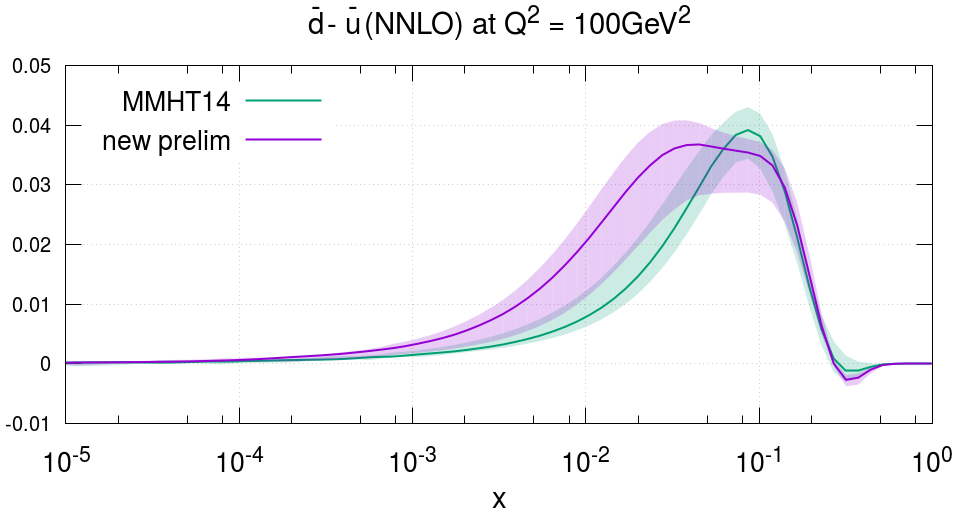}}
\vspace{-0.4cm}
\caption{The NNLO down valence (left) and $\bar d -\bar u$ (right) for the 
fits with extended parameterisation.}
\vspace{-0.2cm}
\label{Fig2} 
\end{figure}

\begin{figure}[]
\centerline{\includegraphics[width=0.5\textwidth]{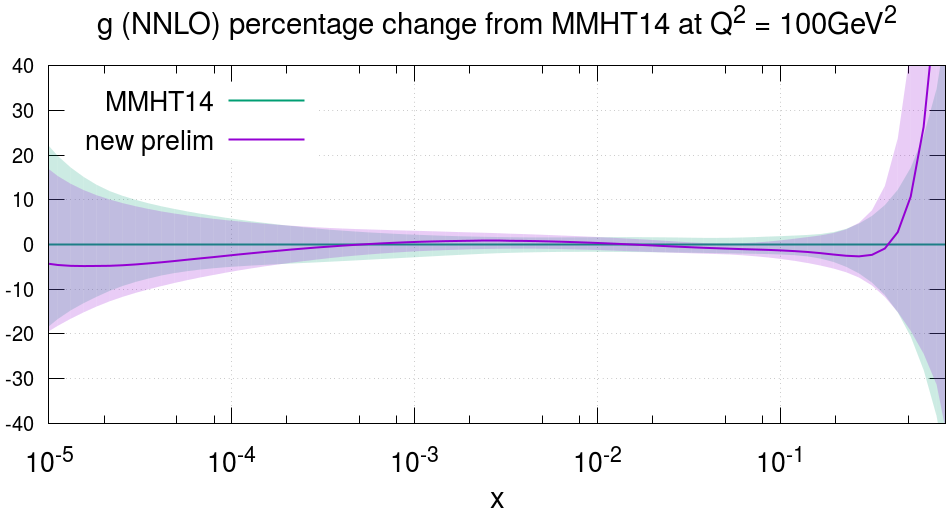}
\includegraphics[width=0.5\textwidth]{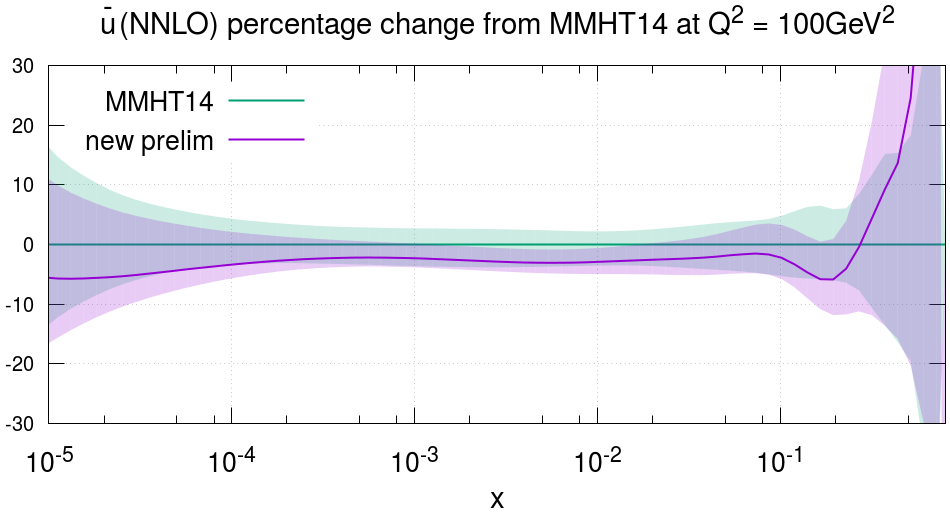}}
\vspace{-0.4cm}
\caption{The NNLO gluon (left) and $\bar u$ (right) for the 
fits with extended parameterisation.}
\vspace{-0.2cm}
\label{Fig3} 
\end{figure}

As well as these main results on the way to presenting the new updated
``MMHT2019'' PDFs, we note some other recent developments. 
In a previous study \cite{Harland-Lang:2017ytb} we found difficulty in fitting 
ATLAS 7~TeV inclusive jet data \cite{Aad:2014vwa}, even with full NNLO
\cite{Currie:2016bfm}. 
To alleviate this we investigated decorrelation of some systematic 
uncertainties between rapidity bins, finding dramatic improvement is we consider 
just two uncertainties, and little sensitivity of the gluon to the decorrelation procedure. This led to more sophisticated proposals for similar 
methods in \cite{Aaboud:2017dvo}. We have also recently tried fitting 
to more than one of the top quark differential distributions in 
\cite{Aad:2015mbv}, 
which is possible due to the 
knowledge of statistical correlations between distributions\cite{ATLAStop},
again using full NNLO \cite{Czakon:2017dip}.
We see a similar picture, i.e. one can fit some individual distributions
well, but not simultaneously. This is due to inconsistencies in the 
required shifts related to correlated systematics (the same result is noted 
and studied in \cite{ATLAStop} and by the CT collaboration). In this case 
the important systematics are all related to Monte Carlo generation, and 
not obviously as well correlated as presented. Significant improvement can be 
achieved on relaxation of correlations. However, in this case the gluon is 
more sensitive to the details of the treatment, and some distributions 
require decorrelation within that particular 
distribution to enable a good fit. More 
details will be presented in the future. 

We also note that we have completed the study of the inclusion of QED effects 
and the photon parton distribution \cite{Harland-Lang:2019pla}. 
We use essentially the PDFs of \cite{Harland-Lang:2016yfn} as a baseline
and use the LUX \cite{Manohar:2016nzj,Manohar:2017eqh} procedure for the 
input photon, making it fully consistent with the MMHT framework, 
including the production of the photon PDF in the neutron. 
We also examine the impact on high-mass Drell-Yan data, noting that 
the photon initiated contributions are sometimes smaller than the QED 
effects on quark evolution. We have 
made grids available, and in particular have separated the elastic 
and inelastic contributions to the photon PDF.

\section*{Acknowledgements}

\vspace{-0.2cm}

LHL thanks the Science and Technology Facilities Council (STFC) for support 
via grant award ST/P004547/1. RST and TC thank the Science and 
Technology Facilities Council (STFC) for support via grant awards ST/P000274/1.
RN thanks the Spreadbury Fund. SB thanks the STFC.


\vspace{-0.2cm}


\begin{thebibliography}{99}


\bibitem{Harland-Lang:2014zoa}
  L.~A.~Harland-Lang, A.~D.~Martin, P.~Motylinski and R.~S.~Thorne,
  Eur.\ Phys.\ J.\ C {\bf 75} (2015) no.5,  204
  doi:10.1140/epjc/s10052-015-3397-6
  [arXiv:1412.3989 [hep-ph]].



\bibitem{Abramowicz:2015mha}
  H.~Abramowicz {\it et al.} [H1 and ZEUS Collaborations],
  Eur.\ Phys.\ J.\ C {\bf 75} (2015) no.12,  580
  doi:10.1140/epjc/s10052-015-3710-4
  [arXiv:1506.06042 [hep-ex]].

\bibitem{Harland-Lang:2016yfn}
  L.~A.~Harland-Lang, A.~D.~Martin, P.~Motylinski and R.~S.~Thorne,
  Eur.\ Phys.\ J.\ C {\bf 76} (2016) no.4,  186
  doi:10.1140/epjc/s10052-016-4020-1
  [arXiv:1601.03413 [hep-ph]].

\bibitem{Thorne:2017aoa}
  R.~S.~Thorne, L.~A.~Harland-Lang and A.~D.~Martin,
  PoS DIS {\bf 2017} (2018) 202
  doi:10.22323/1.297.0202
  [arXiv:1708.00047 [hep-ph]].


\bibitem{Aaboud:2016btc}
  M.~Aaboud {\it et al.} [ATLAS Collaboration],
  Eur.\ Phys.\ J.\ C {\bf 77} (2017) no.6,  367
  doi:10.1140/epjc/s10052-017-4911-9
  [arXiv:1612.03016 [hep-ex]].



\bibitem{Goncharov:2001qe}
  M.~Goncharov {\it et al.} [NuTeV Collaboration],
  Phys.\ Rev.\ D {\bf 64} (2001) 112006
  doi:10.1103/PhysRevD.64.112006
  [hep-ex/0102049].


\bibitem{Berger:2016inr}
  E.~L.~Berger, J.~Gao, C.~S.~Li, Z.~L.~Liu and H.~X.~Zhu,
  Phys.\ Rev.\ Lett.\  {\bf 116} (2016) no.21,  212002
  doi:10.1103/PhysRevLett.116.212002
  [arXiv:1601.05430 [hep-ph]].



\bibitem{Martin:2012da}
  A.~D.~Martin, A.~J.~T.~M.~Mathijssen, W.~J.~Stirling, R.~S.~Thorne, B.~J.~A.~Watt and G.~Watt,
  Eur.\ Phys.\ J.\ C {\bf 73} (2013) no.2,  2318
  doi:10.1140/epjc/s10052-013-2318-9
  [arXiv:1211.1215 [hep-ph]].

\bibitem{Towell:2001nh}
  R.~S.~Towell {\it et al.} [NuSea Collaboration],
  Phys.\ Rev.\ D {\bf 64} (2001) 052002
  doi:10.1103/PhysRevD.64.052002
  [hep-ex/0103030].




\bibitem{Harland-Lang:2017ytb}
  L.~A.~Harland-Lang, A.~D.~Martin and R.~S.~Thorne,
  Eur.\ Phys.\ J.\ C {\bf 78} (2018) no.3,  248
  doi:10.1140/epjc/s10052-018-5710-7
  [arXiv:1711.05757 [hep-ph]].

\bibitem{Aad:2014vwa}
  G.~Aad {\it et al.} [ATLAS Collaboration],
  JHEP {\bf 1502} (2015) 153
   Erratum: [JHEP {\bf 1509} (2015) 141]
  doi:10.1007/JHEP02(2015)153, 10.1007/JHEP09(2015)141
  [arXiv:1410.8857 [hep-ex]].

\bibitem{Currie:2016bfm}
  J.~Currie, E.~W.~N.~Glover and J.~Pires,
  Phys.\ Rev.\ Lett.\  {\bf 118} (2017) no.7,  072002
  doi:10.1103/PhysRevLett.118.072002
  [arXiv:1611.01460 [hep-ph]].

\bibitem{Aaboud:2017dvo}
  M.~Aaboud {\it et al.} [ATLAS Collaboration],
  JHEP {\bf 1709} (2017) 020
  doi:10.1007/JHEP09(2017)020
  [arXiv:1706.03192 [hep-ex]].

\bibitem{Aad:2015mbv}
  G.~Aad {\it et al.} [ATLAS Collaboration],
  Eur.\ Phys.\ J.\ C {\bf 76} (2016) no.10,  538
  doi:10.1140/epjc/s10052-016-4366-4
  [arXiv:1511.04716 [hep-ex]].

\bibitem{ATLAStop}
[ATLAS Collaboration],
ATL-PHYS-PUB-2018-017.



\bibitem{Czakon:2017dip}
  M.~Czakon, D.~Heymes and A.~Mitov,
  arXiv:1704.08551 [hep-ph].


\bibitem{Harland-Lang:2019pla}
  L.~A.~Harland-Lang, A.~D.~Martin, R.~Nathvani and R.~S.~Thorne,
  arXiv:1907.02750 [hep-ph].


\bibitem{Manohar:2016nzj}
  A.~Manohar, P.~Nason, G.~P.~Salam and G.~Zanderighi,
  Phys.\ Rev.\ Lett.\  {\bf 117} (2016) no.24,  242002
  doi:10.1103/PhysRevLett.117.242002
  [arXiv:1607.04266 [hep-ph]].

\bibitem{Manohar:2017eqh}
  A.~V.~Manohar, P.~Nason, G.~P.~Salam and G.~Zanderighi,
  JHEP {\bf 1712} (2017) 046
  doi:10.1007/JHEP12(2017)046
  [arXiv:1708.01256 [hep-ph]].




\end{thebibliography}
\end{document}